# What National Examinations Reforms should be made and how may technology be leveraged?

By Loo Kang WEE Lawrence_Wee@moe.gov.sg Ministry of Education, Singapore

## Abstract
This paper argues for reforming National Examination using a three prong approach and they are 1) monitoring examinations where issues of politics, practicality, security, fairness, accessibility etc are addressed enhancing test items with multimedia (animation and sound ) technologies 2) coordinating playful and stimulating teaching and learning tasks through well managed inter-agencies (Singapore Examinations and Assessment Board SEAB, Ministry of Education etc) implementation and public engagement and 3) portfolio of performance tasks, as an agricultural educational instrument to promote creativity and problems solving, through artifacts of mastery learning and lifelong learning. References are made to CITO[1] based in Netherlands, one of the world's leading testing and assessment companies where the author prototype some demonstration examples using open technology tools such as customized Simple Machines Forum[2] , Open Source Physics tools[3] (Wee & Ning, 2014),  and Blogger[4] etc.

## 1. Introduction

Many may be tempted to ask the following question "How may technology be further leveraged into the National Examinations?" This leads to a bigger question about modernization changes on National Examinations needed, in order to prepare future ready learners of Singapore.

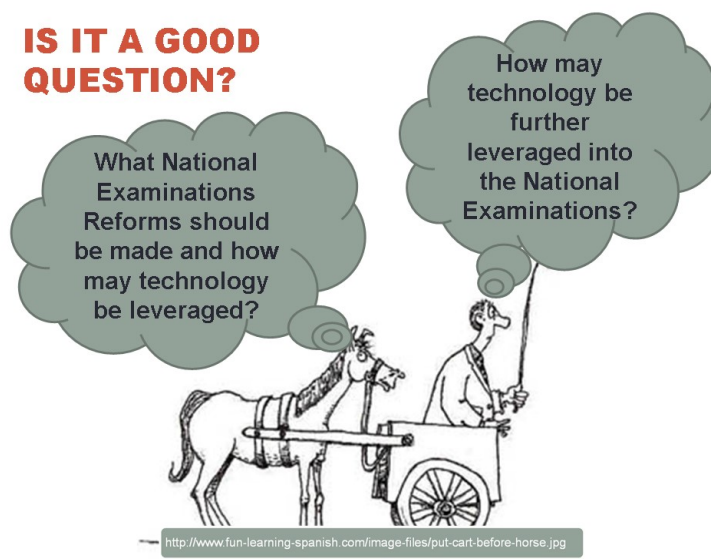

Figure 1.   The key question is on "What national examinations reforms should be made", to nurture futrure ready learners.

---

[1] http://www.cito.com/
[2] http://www.phy.ntnu.edu.tw/ntnujava/index.php
[3] http://iwant2study.org/ospsg/index.php/interactive-resources/physics/01-measurements/5-vernier-caliper
[4] http://weelookang.blogspot.sg/

## 2. Literature

Chakwera, Khembo, & Sireci (2004) provided an objective account on the political, practical, security, psychometric measurement, reality, fairness and access issues when implementing high-stake testing in Malawi, Africa. Baird & Lee‐Kelley (2009) detailed some of lack of manageralism in implementing national examination policy in England, UK and argue for better coordination and more time of various stakeholders to prototype in stages, warning again a hasty national roll-out approach.

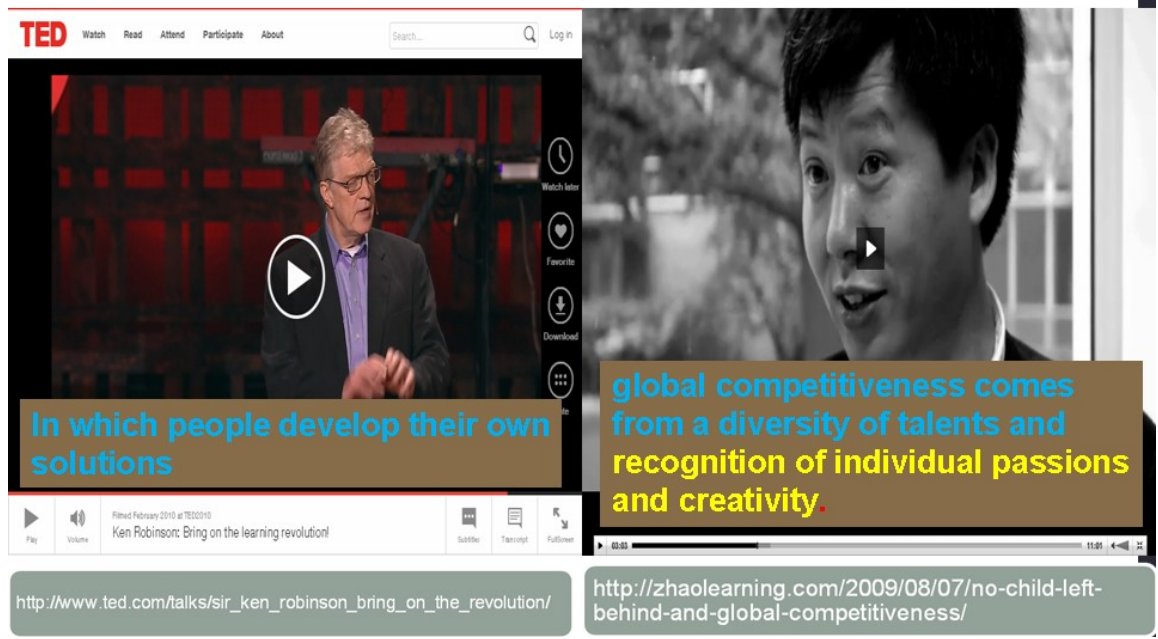

Figure 2. Agricultural education activists such as Ken Robinson (left) and Zhao Yong (right) argue for reforms in education requiring a growth mindset instead of a factory production mindset.

In the field of educational research, prominent scholars like Ken Robinson[5] and Zhao Yong[6] are activists for an agricultural model of education to nurture creative class (Zhao, 2013) of future learners , as oppose to the current factory model of batching students with common core knowledge and skills.

## 3. Possible Solution

In synthesizing the literature on current national examinations and promising future agricultural model of education, one possible solution by CITO, Netherlands, is a 3 prong approach, 1) monitoring examinations, where a variety of issues (Chakwera et al., 2004) are addressed, 2) integration of stimulating teaching and learning tasks, need for solid inter agencies management and coordination (Baird & Lee‐Kelley, 2009), and 3) portfolios as performance of learning, in preparation of an agricultural model of education, promoting learning for life and mastery[7].

---
[5] http://www.ted.com/talks/sir_ken_robinson_bring_on_the_revolution/
[6] http://zhaolearning.com/2009/08/07/no-child-left-behind-and-global-competitiveness/

### i.   Monitoring examinations

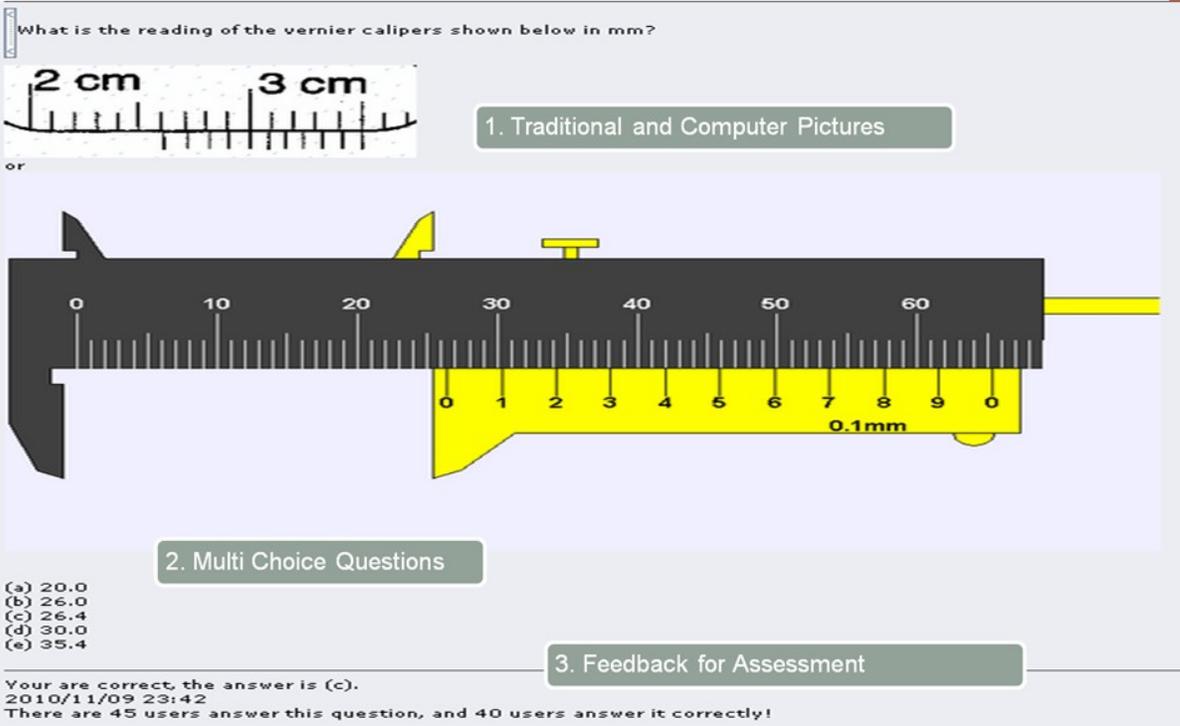

Figure 3.   An example of how a prototype monitoring examination can look like, with traditional and computer pictures, typical multi-choice questions and feedback for assessment.

A prototype multi-choice question[8] could look this, where traditional questions are asked with sound instructional design presenting traditional and computer format together and feedback for assessment for learning. Issues of security and psychometric measurement can be further addressed here through unique logins and authentication.

### ii.   Stimulating Teaching & Learning Tasks + Real Life

As concerted stakeholders of national examinations, stimulating teaching and learning tasks coupled with real equipment (Wee & Ning, 2014) need to be distributed with professional development workshops for teachers. Using resources that are publicly (Wee, Chew, Goh, Tan, & Lee, 2012; Wee, Lee, Chew, Wong, & Tan, 2015) available lower the issues of equity access and practical implementation.

---

[7] http://www.moe.gov.sg/media/speeches/2015/03/06/moe-fy-2015-committee-of-supply-debate-speech-by-minister-minister-heng-swee-keat.php
[8] **http://www.phy.ntnu.edu.tw/ntnujava/index.php?topic=1992.0**

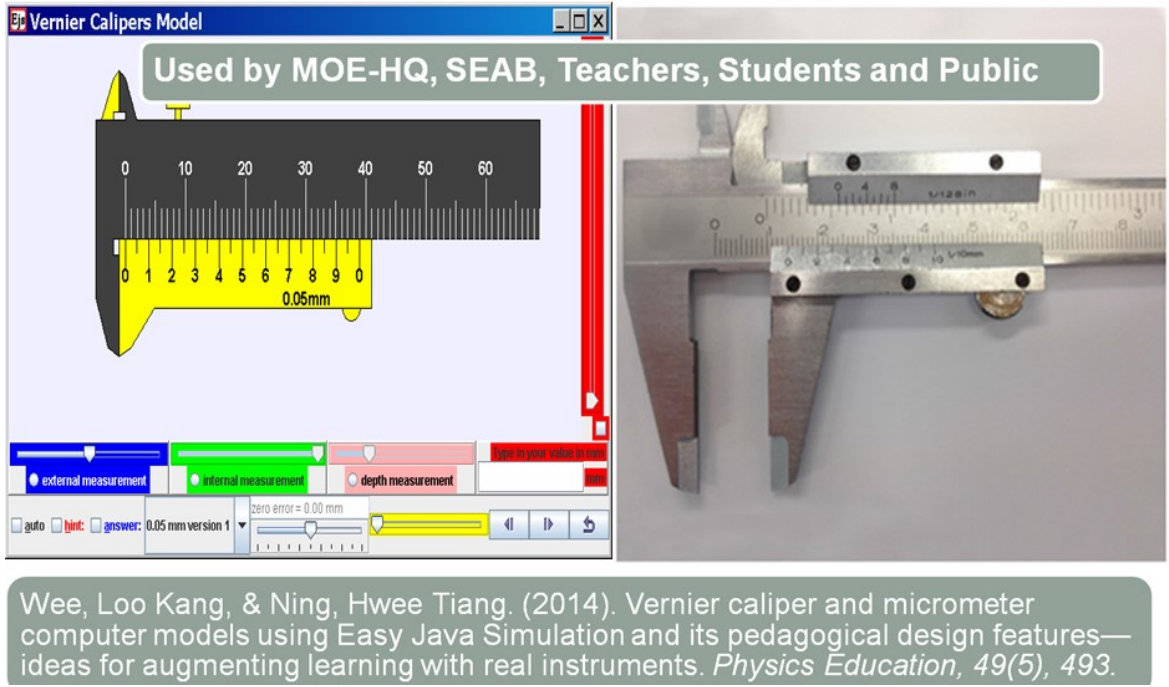

Figure 4. An example of how tapping on open educational resources [9] available on the public internet, lower the issues of equity access and practical implementation.

### iii. Personalized and agricultural Portfolio

As for nurturing future ready learners with personalized and agricultural model of mentoring from teachers, a self-directed system of demonstration of performance tasks (Wee & Leong, 2015) could take any form that the learner's choose, mimicking real life situations facing adulthood career choices.

---

[9] http://iwant2study.org/ospsg/index.php/interactive-resources/physics/01-measurements/5-vernier-caliper

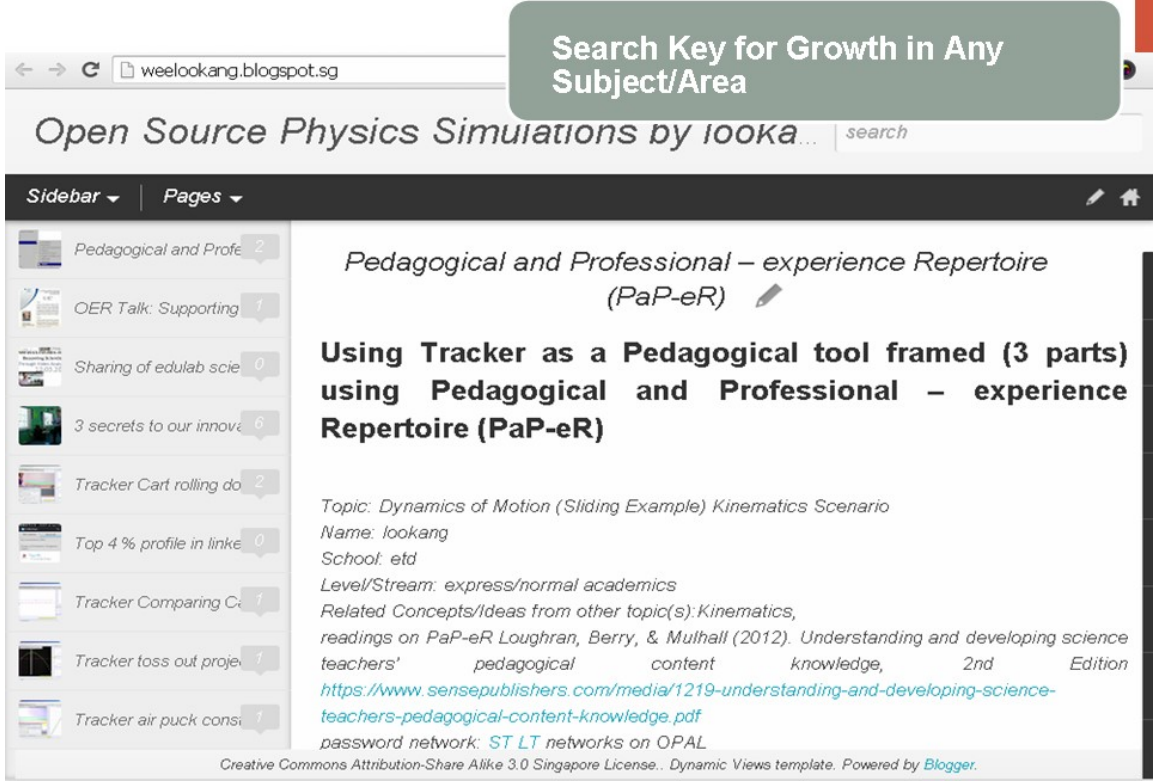

Figure 5. An example of how tapping on technology platform such as Blogger serves as a self-directed system of demonstration of performance tasks portfolio maintained by the learners as mentored by teachers and the public.

## 4. Conclusion

Thus, the table below summarizes the three prong approach 1) monitoring examinations 2) playful and stimulating teaching and learning tasks and 3) portfolio of performance tasks, and the free and open technology tools to be leveraged on for national examinations reforms needed to prepare future ready learners.

| Reforms | Technology |
|---|---|
| Monitoring Exams | Learning Management Space, Big Data Analytics, Adaptive Feedback System, Automated Marking tool |
| Play & Stimulating Teaching & Learning Tasks | Open Source Interactive Tools, Mobile Learning tools, Real Apparatus, Games, Inquiry tools, 1to1 computing |
| Portfolios | Blogs, YouTube, Twitter, Facebook, arXiv, ResearchGate, Google Scholar, LinkedIn |